\documentclass[prd,preprint,showpacs,preprintnumbers,amsmath,amssymb,eqsecnum]{revtex4-1}
\def\t#1{\tilde#1}

\def\d{\delta_{3(1)}}
\def\dd{\delta_{3(2)}}
\def\E{E_{3(0)}}
\def\e{E_{3(1)}}

\usepackage{graphicx}% Include figure files
\usepackage{dcolumn}% Align table columns on decimal point
\usepackage{bm}% bold math
\usepackage{subfigure}
\usepackage{color}
\usepackage{url}

\begin{document}

%\preprint{APS/123-QED}
\preprint{RUP-16-20}
%\preprint{KEK-Cosmo-129}
%\preprint{KEK-TH-1668}
%\preprint{OCU-PHYS 346}
%\preprint{AP-GR 89}

\title{Consistent analytic approach to 
the efficiency of collisional Penrose process}

\author{$^{1}$Tomohiro Harada}%
\email{harada@rikkyo.ac.jp}
\author{$^{1}$Kota Ogasawara}
%\email{??@??}
\author{$^{2}$Umpei Miyamoto}
%\email{??@??}
\affiliation{$^{1}$Department of Physics, Rikkyo University, Toshima,
Tokyo 171-8501, Japan}
\affiliation{$^{2}$RECCS, Akita Prefectural University, Akita 015-0055, Japan}
\date{\today}% It is always \today, today,
\begin{abstract}

We propose a consistent analytic approach to 
the efficiency of collisional Penrose process 
in the vicinity of a maximally rotating Kerr black hole.
We focus on a collision with 
arbitrarily high center-of-mass energy, which occurs 
if either of the colliding particles
 has its angular momentum fine-tuned to the critical value to enter the horizon.
We show that 
if the fine-tuned particle is ingoing on the collision, 
the upper limit of the efficiency
is $(2+\sqrt{3})(2-\sqrt{2})\simeq 2.186$, while 
if the fine-tuned particle is bounced back before the collision, 
the upper limit is $(2+\sqrt{3})^{2}\simeq 13.93$.
Despite earlier claims, the former can be attained 
for inverse Compton scattering if the fine-tuned particle  
is massive and starts 
at rest at infinity, 
while the latter can be attained for various particle reactions, such as 
inverse Compton scattering and pair annihilation, if 
the fine-tuned particle is either massless or highly 
relativistic at infinity.
We discuss the difference between the present and earlier analyses.
\end{abstract}


\maketitle

%\tableofcontents

%\newpage

\section{Introduction}
If two particles collide in the ergosphere of a rotating black hole,
a product particle can escape to infinity with energy greater than the 
total energy of the incident particles. This is called the collisional
Penrose process, and 
was pioneered by Piran et al.~\cite{PSK:1975,Piran:1977dm}. 
The study of this phenomenon has recently been revived since 
Ba\~{n}ados, Silk and West (BSW)~\cite{Banados:2009pr} 
revealed that maximally rotating black holes can accelerate 
particles to arbitrarily high energy if their angular momenta are fine-tuned.
More precisely, if two particles start at rest 
at infinity and if the angular momentum of either of the two 
is fine-tuned to the threshold value to enter the horizon, 
the center-of-mass energy of the 
two colliding particles can be 
arbitrarily high.

It is a fundamental question whether particle acceleration by a 
rotating black hole has anything to do with energy extraction 
from the rotating black hole. 
Although it was claimed~\cite{Jacobson:2009zg} 
that there is no energy extraction for the 
BSW process of two equal masses, 
where both the fine-tuned particle and the generic particle
are ingoing on the collision, 
it was later shown that the energy 
extraction efficiency can reach $\simeq 1.3$ for the BSW process
if the masses of the product particles are not equal to the 
incident particles~\cite{Bejger:2012yb,Harada:2012ap}.
In Ref.~\cite{Harada:2012ap}, 
the authors including two of the present ones also 
claimed that $\simeq 1.4$ is a common upper limit 
for any particle reactions. More recently,
Schnittman~\cite{Schnittman:2014zsa} 
has shown that the upper limit 
can reach $\simeq 14$, which is analytically given by $(2+\sqrt{3})^{2}$~\cite{Leiderschneider:2015kwa,Ogasawara:2015umo}, if the collision 
occurs immediately after the ingoing fine-tuned 
particle is bounced back
outwardly near the horizon by a potential barrier. 
In Ref.~\cite{Ogasawara:2015umo}, 
the present authors discussed that the process of such high efficiency 
would require heavy particle production. 
Berti et al.~\cite{Berti:2014lva} pointed out 
that the efficiency can be arbitrarily
high if we allow a particle with the subcritical value of 
angular momentum to start with an outgoing initial velocity 
in the vicinity of a black hole, although the physical motivation 
of such a particle is controversial~\cite{Leiderschneider:2015ika}.

In the current article, we propose a consistent analytic 
approach and show that the assumption made in the 
previous analytic analyses 
in Refs.~\cite{Harada:2012ap,Ogasawara:2015umo} is too restricted.
Under a physically reasonable assumption, we show that  
the upper limit for the BSW process is given by 
$(2+\sqrt{3})(2-\sqrt{2})\simeq 2.186$, which can be attained 
for inverse Compton scattering if the incident fine-tuned
particle is massive and starts at rest at infinity. We also show that 
the upper limit for Schnittman's process is given by
$(2+\sqrt{3})^{2}\simeq 13.93$, which can be attained 
for various particle reactions if the incident fine-tuned
particle is massless or starts with a highly relativistic velocity
at infinity.

\section{Analytic approach}
We consider the reaction of incident particles 1 and 2 
colliding near the horizon
to two product particles 3 and 4 in the equatorial plane of a
Kerr black hole of mass $M$ and spin $a$.
The energy and angular momentum conservation yields
\begin{equation}
 E_{1}+E_{2}=E_{3}+E_{4},\quad
  L_{1}+L_{2}=L_{3}+L_{4}.
\label{eq:ELconservation}
\end{equation}
The conservation of the total radial momentum immediately before and
after the reaction is given
by 
\begin{equation}
 p^{r}_{1}+p^{r}_{2}=p^{r}_{3}+p^{r}_{4}
\label{eq:pr}
\end{equation} 
at the collision point $r=r_{c}$.
The radial momentum $p^{r}$ of the particle is given by
\begin{equation}
 p^{r}=\sigma \sqrt{-2V(r)},
\end{equation}
where $\sigma=\pm 1$ and 
\begin{equation}
 V(r)=-\frac{Mm^{2}}{r}+\frac{L^{2}-a^{2}(E^{2}-m^{2})}{2r^{2}}
-\frac{M(L-aE)^{2}}{r^{3}}-\frac{E^{2}-m^{2}}{2}.
\end{equation}
Hereafter, we concentrate on maximally rotating black
holes, i.e., $a=M$ for simplicity. For this case, $L=2ME$ gives the
critical value of angular momentum for an ingoing particle 
to enter the horizon from outside.

There are 3 parameters $(E,L,m)$ for each of the
particles except for $\sigma$. 
We have 3 equations and so $3\times 4-3=9$ degrees of freedom
are remaining except for $\sigma_{i}$ ($i=1,2,3,4$). 
If we specify the species of the incident 
particles, we can fix $m_{1}$ and $m_{2}$.
Thus, $9-2=7$ degrees of freedom are remaining.
If we specify the remaining parameters $E_{1}$, $L_{1}$, $E_{2}$ and 
$L_{2}$ for the incident particles, we have $7-4=3$ degrees of freedom
remaining.
Moreover, we can fix $m_{3}$ and $m_{4}$ for a particle reaction we know.
Thus, only one degree of freedom is remaining.
We can take $\delta$ for an escaping particle as this, where $\delta$ 
parameterizes the ratio of $L$ and $E$, i.e., 
\begin{equation}
L=(2+\delta)ME.
\end{equation} 
We should note that $\delta$ essentially determines the orbit of
the particle and, hence, whether it can escape to infinity or not.
If $0<\delta<\delta_{\rm max}(r_{c})$ and $E\ge m$, the particle 
will escape to infinity whether it is
initially outgoing or ingoing, while if $\delta_{\rm min}(r_{c})<\delta<0$, it must
start with an outgoing initial velocity, where 
$\delta_{\rm max}(r)$ and 
$\delta_{\rm min}(r)$ are determined by the turning 
point condition $V(r)=0$. 
This can be seen in Fig. 1 of Ref.~\cite{Ogasawara:2015umo}.
Clearly, we cannot control $\delta$, which corresponds to the direction of
the initial velocities of the product particles. 
We take the energy and angular momentum of 
the escaping particle as functions of those of the incident particles,
the collision point $r_{c}$ and $\delta$. 
That is, identifying particle 3 with that of the escape to infinity,
we have
\begin{eqnarray}
 E_{3}&=&E_{3}(E_{1},L_{1},m_{1},\sigma_{1c};E_{2},L_{2},m_{2},\sigma_{2c};m_{3},
  \sigma_{3c}; m_{4},\sigma_{4c};\delta_{3};r_{c}), \\
 L_{3}&=&L_{3}(E_{1},L_{1},m_{1},\sigma_{1c};E_{2},L_{2},m_{2},\sigma_{2c};m_{3},\sigma_{3c};m_{4},\sigma_{4c};\delta_{3};r_{c}),
\end{eqnarray}
where $\sigma_{ic}$ ($i=1,2,3,4$) are the values of $\sigma_{i}$ 
immediately before and after the collision. The energy extraction 
efficiency $\eta$ is defined as $\eta:=E_{3}/(E_{1}+E_{2})$.

To investigate the collisional Penrose process with 
arbitrarily high center-of-mass energy,
we assume that particles 1 and 2 are critical and subcritical, respectively; 
i.e., $\tilde{L}_{1}=2E_{1}$ and $\tilde{L}_{2}<2E_{2}$, where 
$\tilde{L}:=L/M$ and for which 
the center-of-mass energy behaves as
$E_{\rm cm}\propto (r-r_{c})^{-1/2}$ for
$0<r-r_{c}\ll M $~\cite{Grib:2010zs,Harada:2010yv}.
We assume $\sigma_{2c}=-1$ and $\sigma_{4c}=-1$ on collision 
to specify the process.
To go further, we express $p^{t}$ and $p^{r}$ in terms of $\epsilon$, where 
$r=M/(1-\epsilon)$. 
We note that for fixed $E$ and $L$, $p^{t}$ is given by 
\begin{equation}
 p^{t}=\frac{1}{\epsilon^{2}}[2(2E-\tilde{L})+2(-4E+3\tilde{L})\epsilon+(7E-6\tilde{L})\epsilon^{2}+2(-E+\tilde{L})\epsilon^{3}],
\end{equation} 
while $(p^{r})^{2}$ is given by 
\begin{eqnarray}
 (p^{r})^{2}=[(2E-\tilde{L})-2(E-\tilde{L})\epsilon]^{2}
-[m^{2}-(E-\tilde{L})(3E-\tilde{L})]\epsilon^{2}
-2(E-\tilde{L})^{2}\epsilon^{3}.
\end{eqnarray}

Denoting $r_{c}=M/(1-\epsilon_{c})$, we can find that
Eqs.~(\ref{eq:ELconservation}) and (\ref{eq:pr})
imply that $\sigma_{3c}<0$ or
$2E_{3}-\tilde{L}_{3}=O(\epsilon_{c})$. In the former case, 
we can expect particle 3 to eventually escape to infinity only if
$2E_{3}-\tilde{L}_{3} < 0$, while the forward-in-time condition $p^{t}>0$
implies $2E_{3}-\tilde{L}_{3}>(4E_{3}-3\tilde{L}_{3})\epsilon_{c} +O(\epsilon_{c}^{2})$.
Therefore, whether $\sigma_{3c}=1$ or $-1$, 
we conclude $2E_{3}-\tilde{L}_{3}=O(\epsilon_{c})$, so that particle 3
must be near-critical.

Since $E_{3}$ and $\tilde{L}_{3}$ are functions of the radius of the
collision point $r_{c}$, we can assume that $E_{3}$ and $\tilde{L}_{3}$ are 
expandable in terms of $\epsilon_{c}$, i.e., 
\begin{eqnarray}
 E_{3}&=&E_{3(0)}+E_{3(1)}\epsilon_{c}+E_{3(2)}\epsilon_{c}^{2}+\ldots, \\
 \tilde{L}_{3}&=&\tilde{L}_{3(0)}+\tilde{L}_{3(1)}\epsilon_{c}+\tilde{L}_{3(2)}\epsilon_{c}^{2}+\ldots .
\end{eqnarray}
We expand $E_{i}$ and $\tilde{L}_{i}$ in terms of
$\epsilon_{c}$ for the 
product particles $i=3,4$, but not for incident
particles $i=1,2$. This looks asymmetric but is suitable for the present 
physical setting.
Equivalently, instead of $\tilde{L}_{3}$, it is more convenient to
expand $\delta_{3}$
as follows:
\begin{equation}
 \delta_{3}=\delta_{3(1)}\epsilon_{c}+\delta_{3(2)}\epsilon_{c}^{2}
  +\ldots .
\end{equation}  
If particle 3 is ingoing immediately after the collision
with $E_{3}\ge m_{3}$ and to
escape to infinity, it must
be bounced back by a potential barrier inside the collision point, 
for which 
\begin{eqnarray}
0<\d\leq\delta_{(1),{\rm max}},~~\delta_{(1),{\rm max}}:=\frac{2\E-\sqrt{\E^2+m^2_3}}{E_{3(0)}}.
\end{eqnarray}
If particle 3 is outgoing immediately after the collision
with $E_{3}\ge m_{3}$ and to escape to infinity, 
it must not encounter a potential
barrier outside the collision point and this is guaranteed for a
near-critical particle.

Since we have already seen the terms of $O(1)$ in Eq.~(\ref{eq:pr}), we
proceed to the terms of $O(\epsilon_{c})$ in the same equation. Together
with Eq.~(\ref{eq:ELconservation}), we obtain
\begin{equation}
A-\E(2-\d)=\sigma_{3c}\sqrt{\E^2(3-\d)(1-\d)-m^2_3},
\label{eq:Oepsilon}
\end{equation}
where 
$A:=2E_1+\sigma_{1c}\sqrt{3E^2_1-m^2_1}>0$.
Squaring the both sides of Eq.~(\ref{eq:Oepsilon}), we find 
\begin{eqnarray}
2-\d=\frac{A^2+\E^2+m^2_3}{2A\E}.
\label{eq:2-d}
\end{eqnarray}
Substituting the above into the left-hand side of Eq.~(\ref{eq:Oepsilon}),
we find 
\begin{eqnarray}
A-\frac{\E^2+m^2_3}{A}=2\sigma_{3c}\sqrt{\E^2(3-\d)(1-\d)-m^2_3}.
\end{eqnarray}
For $\sigma_{3c}=1$, we immediately find 
\begin{equation}
 E_{3(0)}\le \lambda_{0},~~\lambda_{0}:=\sqrt{A^{2}-m_{3}^{2}}.
\end{equation}
For $\sigma_{3c}=-1$, where $\delta_{3(1)}\ge  0$ must be satisfied for 
particle 3 to escape to infinity,
 Eq.~(\ref{eq:2-d}) yields
\begin{eqnarray}
\E^2-4A\E+A^2+m^2_3\leq0.
\end{eqnarray}
Thus, we find 
\begin{eqnarray}
\lambda_-\leq\E\leq\lambda_+\;\;,\;\;\lambda_\pm:=2A\pm\sqrt{3A^2-m^2_3},
\end{eqnarray}
where $E_{3(0)}=\lambda_{+}$ is realized only for $\d=0$.

Therefore, for given $E_{1}$, $E_{2}$ and $m_{i}$
($i=1,2,3,4$), we find for $\sigma_{3c}=1$
\begin{equation}
 \eta_{\rm
  max}=\frac{\sqrt{(2E_{1}+\sigma_{1c}\sqrt{3E_{1}^{2}-m_{1}^{2}})^{2}-m_{3}^{2}}}{E_{1}+E_{2}}, 
\label{eq:eta_max_1}
\end{equation}
while for $\sigma_{3c}=-1$
\begin{equation}
 \eta_{\rm max}=\frac{2(2E_{1}+\sigma_{1c}\sqrt{3E_{1}^{2}-m_{1}^{2}})+
\sqrt{3(2E_{1}+\sigma_{1c}\sqrt{3E_{1}^{2}-m_{1}^{2}})^{2}-m_{3}^{2}}}{E_{1}+E_{2}},
\label{eq:eta_max_-1}
\end{equation}
where $\sigma_{1c}=-1$ and $1$ correspond to the BSW and Schnittman
processes, respectively. We can see that the upper limit for
$\sigma_{3c}=-1$ is always greater than that for $\sigma_{3c}=1$.

The terms of $O(\epsilon_{c}^2)$ in Eq.~(\ref{eq:pr}) yield
\begin{eqnarray}
&&\sigma_{1c}\frac{E^2_1}{\sqrt{3E^2_1-m^2_1}}
+\frac{(3E_2-\tilde L_2)(E_2-\tilde L_2)-m^2_2}{2(2E_2-\tilde L_2)}
\nonumber\\&&=-\sigma_{3c}
\frac{\E^2[-1+(2-\d)(2\d-\dd)]+\E\e(3-\d)(1-\d)}{\sqrt{\E^2(3-\d)(1-\d)-m^2_3}}
\nonumber\\&&~~~+
\frac{2E_2-\tilde
L_2}{2}-\E(2\d-\dd)-\e(2-\d)-\frac{(E_1+E_2-\E)^2+m^2_4}{2(2E_2-\tilde
L_2)}.~~~~~~~
\label{eq:epsilon_c^2}
\end{eqnarray}
If we fix $E_{3(0)}$ and $\delta_{3(1)}$, 
we find the relation
between $E_{3(1)}$ and $\delta_{3(2)}$ from the above
equation. 
Since both $E_{3(1)}$ and
$\delta_{3(2)}$ appear only linearly, we can always solve the above
equation for $E_{3(1)}$ in terms of $\delta_{3(2)}$.
We do not obtain any additional condition 
to the lower-order terms.

Here we still take $m_{i}$ ($i=1,2,3,4$) as fixed parameters but 
$E_{1}$ and $E_{2}$ as free ones in the ranges $E_{1}\ge m_{1}$ 
and $E_{2}\ge m_{2}$, respectively.
For the BSW process with $m_{1}>0$ and $m_{2}>0$, 
the maximum efficiency is attained 
for $E_{1}=m_{1}$ and $E_{2}=m_{2}$ as 
we can see from Eqs.~(\ref{eq:eta_max_1})
and (\ref{eq:eta_max_-1}) with $\sigma_{1c}=-1$.
The upper limit of the efficiency for $\sigma_{3c}=1$ is given by 
\begin{equation}
 \eta_{\rm max}=\frac{\sqrt{(2-\sqrt{2})^{2}m_{1}^{2}-m_{3}^{2}}}{m_{1}+m_{2}}, 
\end{equation}
which is less than unity, 
while for $\sigma_{3c}=-1$ it is given by
\begin{equation}
 \eta_{\rm max}=\frac{2(2-\sqrt{2})m_{1}+\sqrt{3(2-\sqrt{2})^{2}m_{1}^{2}-m_{3}^{2}}}{m_{1}+m_{2}}. 
\end{equation}
The above expression applies also for $m_{2}=0$.
For perfectly elastic collision of equal masses, we find
$\eta_{\rm max}=(7-4\sqrt{2})/2<1$ even for $\sigma_{3c}=-1$. 
But if we change the masses of the particles,
the upper limit can be greater than unity. For pair annihilation, 
$\eta_{\rm max}=(2-\sqrt{2})(2+\sqrt{3})/2\simeq 1.093$ for $\sigma_{3c}=-1$,
which agrees very well with the numerical result
given in Ref.~\cite{Bejger:2012yb}. 
For inverse Compton with $m_{2}=m_{3}=0$, 
$\eta_{\rm max}=(2-\sqrt{2})(2+\sqrt{3})\simeq 2.186$, which
is realized for $E_{1}=m_{1}\gg E_{2}$. This is the maximum
upper limit of the efficiency for the BSW process. 
For the BSW process with $m_{1}=0$, 
the upper limit of the efficiency for $\sigma_{3c}=1$ is given by 
$\eta_{\rm max}=2-\sqrt{3}$, while for $\sigma_{3c}=-1$ the upper limit 
is given by
$\eta_{\rm max}=1$. These upper limits are realized for $E_{1}\gg
\mbox{max}(E_{2},m_{3})$. In particular, 
for inverse Compton with $m_{1}=m_{3}=0$,
we can see $\eta_{\rm max}=1$.

For Schnittman's process ($\sigma_{1c}=1$), the situation is very
different and much simpler. 
The upper limit can be attained for $E_{1}\gg \max(m_{1},m_{3})$
and $E_{1}\gg E_{2}$ as we can see from 
Eqs.~(\ref{eq:eta_max_1}) and
(\ref{eq:eta_max_-1}) with $\sigma_{1c}=1$. In this case, we find $\eta_{\rm max}=2+\sqrt{3}\simeq 3.732$ and 
$(2+\sqrt{3})^{2}\simeq 13.93$ for $\sigma_{3c}=1$ and $-1$, 
respectively. The latter is a universal upper limit irrespective of 
the details of the particle reaction or the masses of the particles.

\section{Discussion and summary}
Here, we compare the current analysis with earlier 
ones~\cite{Harada:2012ap,Ogasawara:2015umo}.
For example, if we fix the parameters such as 
$\sigma_{1c}=1$, $\sigma_{3c}=-1$ and $\d=0$, i.e., those for 
Schnittman's process, and solve
Eq.~(\ref{eq:epsilon_c^2}) for $m^2_4$, we find 
\begin{eqnarray}
m^2_4=2(2E_2-\t L_2)F+E^2_2+m^2_2-(E_1+E_2-\E)^2,
\label{m^2_4=}
\end{eqnarray}
where 
\begin{eqnarray}
F&=&-\frac{E^2_1}{\sqrt{3E^2_1-m^2_1}}-\frac{\E^2+\E\dd(2\E-\sqrt{3\E^2-m^2_3})}{\sqrt{3\E^2-m^2_3}}
 \nonumber \\
%\nonumber\\&&~~~~~~~
& & \quad +\frac{\e(3\E-2\sqrt{3\E^2-m^2_3})}{\sqrt{3\E^2-m^2_3}}.
\label{F}
\end{eqnarray}
If we assume $E_{3(1)}\ge  0$ and $\delta_{3(2)}\ge 0$, we can conclude that
$F\le 0$ and, hence, $E_{2}$ is bounded from below. This places a strong limit on the upper limit. 
However, if we allow $E_{3(1)}$ to be negative, the sign of $F$ is
indefinite and 
there is no constraint on $E_{2}$ and $m_{2}$ except for
the initial assumption $E_{2}\ge m_{2}$. In Refs.~\cite{Harada:2012ap,Ogasawara:2015umo},
the authors assumed that $E_{3}$ does not depend on
$\epsilon_{c}$, which is equivalent to
$E_{3(1)}=0$ in the current analysis. 
This assumption is apparently too restricted. 
In Ref.~\cite{Harada:2012ap}, it gives a rather smaller
upper limit, while in Ref.~\cite{Ogasawara:2015umo}, 
the possibility of heavy particle production has to be discussed.
The current analysis implies that 
$E_{3(1)}$ is negative for the upper limit case and suggests that 
the upper limit is realized only 
in the near-horizon limit of Schnittman's process.

Bejger et al.~\cite{Bejger:2012yb} numerically demonstrated that
for the BSW process of pair annihilation  
the upper limit can reach 1.295, which 
is greater than our near-horizon limit 1.093. 
Note that their value is not realized in a simple near-horizon limit.
We believe that their limit value 1.295 will be 
obtained if the angular momentum of 
particle 2 is also fine-tuned so that $2E_{2}-\tilde{L}_{2}=O(\epsilon_{c})$,
i.e., particle 2 is near-critical.
It would be interesting to pursue this direction further in the 
generalization of the present analytic approach.

In summary, we propose a consistent analytic approach to the efficiency 
of the collisional Penrose process with the 
expansion in powers of the small parameter $\epsilon_{c}$, 
which parameterizes the closeness of the collision point to the horizon.
According to this systematic approach, we find that the upper limits of 
the collisional Penrose process restricted in the equatorial plane 
near the horizon are given by 
$(2+\sqrt{3})(2-\sqrt{2})\simeq 2.186$ and 
$(2+\sqrt{3})^{2}\simeq 13.93$ for the BSW and Schnittman processes, 
respectively. The former is realized for inverse Compton scattering, 
while the latter can be universally attained for various reaction of
particles. In spite of the earlier 
claims~\cite{Harada:2012ap,Ogasawara:2015umo}, these upper 
limits can be realized for standard particle reactions such as 
inverse Compton scattering and pair annihilation.

\acknowledgments

The authors would like to thank E. Gourgoulhon, H. Ishihara, 
T. Igata, 
K. Nakao,
M. Patil,
K. Tanabe, 
and H. Yoshino for fruitful discussion. 
This work was supported by JSPS KAKENHI Grants No. 26400282 (T.H.)
and 15K05086 (U.M.).

\end{document}